# Covert Surveillance in Smart Devices: A SCOUR Framework Analysis of Youth Privacy Implications

Austin Shouli
Computer *Science Department*
*Vancouver Island University*
Nanaimo, BC, Canada
austin.shouli@viu.ca

Yulia Bobkova
Computer *Science Department*
*Vancouver Island University*
Nanaimo, BC, Canada
yulia.bobkova@viu.ca

Ajay Kumar Shrestha
Computer *Science Department*
*Vancouver Island University*
Nanaimo, BC, Canada
ajay.shrestha@viu.ca

*Abstract—* **This paper investigates how smart devices covertly capture private conversations and discusses in more in-depth the implications of this for youth privacy. Using a structured review guided by the PRISMA methodology, the analysis focuses on privacy concerns, data capture methods, data storage and sharing practices, and proposed technical mitigations. To structure and synthesize findings, we introduce the SCOUR framework, encompassing Surveillance mechanisms, Consent and awareness, Operational data flow, Usage and exploitation, and Regulatory and technical safeguards. Findings reveal that smart devices have been covertly capturing personal data, especially with smart toys and voice-activated smart gadgets built for youth. These issues are worsened by unclear data collection practices and insufficient transparency in smart device applications. Balancing privacy and utility in smart devices is crucial, as youth are becoming more aware of privacy breaches and value their personal data more. Strategies to improve regulatory and technical safeguards are also provided. The review identifies research gaps and suggests future directions. The limitations of this literature review are also explained. The findings have significant implications for policy development and the transparency of data collection for smart devices.**

## I. INTRODUCTION

Privacy is shorthand for 'breathing space' that encourages self-expression and gives us the freedom to do and be as we like, without the fear of public judgment [1]. The digital technologies that have revolutionized our lives have also created a detailed shadow of our lives on record, available as data [2]. Smart voice assistant (VA) use has expanded from personal uses in the home to new applications in customer services, healthcare, e-government, and educational spaces, raising questions from groups like the American Civil Liberties Union (ACLU), among others, about the data privacy implications of these technologies in public and shared spaces [3]. Smart speakers with voice assistants, like Amazon Echo and Google Home, provide benefits and convenience but also raise privacy concerns due to their continuously listening microphones[4]. Smart speakers often listen continuously to their surroundings, enabling them to respond to a wake word. The privacy implications of this have been the focus of much public debate, especially in the intimacy of personal homes [3]. Toys labelled as "smart" are being sold to many consumers worldwide, but it is often unclear what privacy implications accompany these toys. Children using these toys may be too young to understand the privacy concerns behind them. Certain devices capture the everyday activities of children, along with their biometric data (voice, fingerprints, etc.), which are tracked, recorded and analyzed [5]. Furthermore, smart toys could potentially process the data of the people around the environment of child. As the rate of technological adoption accelerates, it is important to consider the right to preserve a certain amount of privacy and independence in the home [4]. Studies show that humans often have poor mental models regarding always-listening assistants, and are often ill-equipped to grasp the large amount of private information being captured [5]. Although both users and peripheral non-users are typically cognizant of some potential privacy risks, many users have decided to adopt the devices nonetheless, trading convenience for privacy [6].

Given these challenges and the complex landscape of privacy in smart devices, a systematic literature review is essential to comprehensively understand the perceptions, concerns, and expectations of users, with an emphasis on youth regarding their privacy. This systematic literature review synthesizes findings from various studies to answer key questions about privacy concerns, data-sharing practices, methods of capturing data, prevention and precaution to give the user control over personal data, with a particular focus on Canadian regulation and young users. This paper emphasizes the need for transparent smart device manufacturing, data practices, and design privacy notices. Users' intention to use new technology is affected by trust, and technology adoption is related to trust concerns [7]. There is a risk that the growing awareness of digital exposure may negatively affect people's experience of digital technologies, decrease their trust in online services, or even completely deter them from using digital technology [8]. To build trust, it is essential to ensure clear communication about data practices, methods of capturing information, and storage of data [9]. Additionally, transparency expectations vary across contexts, with higher demands for privacy in smart home settings.

The remainder of the paper is structured as follows: Section II provides the background and outlines the context of the study. Section III details the methods used in the study. Results are presented in Section IV, with the discussion and limitations following in Section V. Finally, Section VI concludes the paper.

## II. BACKGROUND

This background provides the reader with an overview of the technologies and concepts necessary for understanding this systematic literature review. This review aims to contribute to the existing body of work regarding the risks posed by smart devices, including the risk of surveillance and general privacy risks posed to users.

*A. Smart and AI-enabled Devices*

Many terms are used interchangeably to refer to the range of privacy-impacting consumer devices that make up the Internet of Things (IoT). Most pertinent to this study are smart speakers. These are consumer electronic devices that are placed in the home, consisting of microphones, speakers, and networked computers that typically communicate with remote servers for processing. Users are able to interact with the device through spoken commands, which the device interprets and responds to using AI models. Common uses are for playing music, setting times, or asking questions. Popular products in this category include Amazon's Echo and Google's Home and Nest speakers. Another relevant device category is smartphones, such as Apple's iPhone and the accompanying Siri voice assistant. Similar to smart speakers, to respond to a user's vocalized request, these devices employ an "always on" approach, using microphones that are continually listening to their surroundings to activate them with a wake word [10]. This poses various privacy concerns, including the risk of inadvertent activation and audio data storage/collection protocols, for the device's user and others in proximity to the device. Smart recording devices and applications can also be found in children's toys, automobiles, headphones, and other devices. Additional terminology is used throughout the literature to refer to a broad range of smart, connected devices, including ambient intelligence, internet of things (IoT), and smart homes, and variations on the term (VA) and intelligent personal assistant (IPA) [11], [12], [13], [14], [15].

*B. Privacy Concerns Across Smart Devices*

Due to the nature of always-on devices, there is a heightened risk of inadvertent and malicious surveillance. Research has shown that Amazon employees manually review material recorded by their Echo devices to improve their services. This behavior is enabled by default, requiring a user to manually opt out [10]. There are serious concerns regarding the conditions that allow for a smart device to be activated, how inadvertent activations are handled, and the data storage and use policies for collected auditory data [16]. These concerns are amplified as technology progresses and industry increasingly shifts towards the use of an always-on, passively listening approach [11]. These concerns are exacerbated by a general lack of public understanding regarding the basic functionality and underlying mechanisms of these devices [13], [17]. Additionally, cybersecurity concerns arise from the data collection, use and storage of data collected by smart devices [18]. This cybersecurity risk increases with the ecosystem of third-party applications and integrations within the smart device ecosystem [11].

*C. Existing Regulatory Frameworks*

Canada's Personal Information Protection and Electronic Documents Act (PIPEDA) does provide regulatory guidance on the collection and use of personal information, defined as "information about an identifiable individual" [19]. These restrictions apply to businesses operating in Canada, providing a level of protection for consumer data collected and processed by smart devices. Additionally, various provincial privacy acts exist, which may impose additional requirements at a provincial level. Globally, there are similar frameworks such as Europe's General Data Protection Regulation (GDPR), which provides a wide range of data and privacy protections for individuals, including the right to erase personal data [20]. While the United States lacks a comprehensive, federal privacy framework, always-on devices may run contrary to wiretap laws in various states [11].

### III. METHODOLOGY

We employed a systematic literature review approach based on the Preferred Reporting Items for Systematic reviews and Meta-Analyses (PRISMA) framework, as shown in Fig. 1, to identify existing works pertaining to our research question and evaluate the current body of research [21], [22]. We then proceeded using the SCOUR framework, as outlined in Table 1, to categorically sort the papers identified and to extract key information from each paper. The result of this approach was to understand prior research that has been conducted, identify key concepts and themes, and provide a foundation of knowledge for future research to be conducted in this area.

*A. Research Question*

The primary question we set out to answer was: *Can smart/AI-enabled devices, applications, intentionally and surreptitiously record private conversations? If so, what evidence exists of this practice?*

To gain a broader context of this research question, secondary considerations included:
- What mechanisms allow such recordings to occur?
- How is the collected data used, stored, and shared?
- Are there documented cases where companies or third parties have leveraged such recordings?
- What are the technical, policy, and legal limitations regarding surreptitious recording?
- What preventive measures can individuals and regulators take to mitigate unintended recordings?

*B. Literature Collection*

A systematic review approach was applied to identify relevant literature. Our team used Python scripts to scrape papers from the Google Scholar and Crossref databases using their respective APIs and a set of relevant keywords combined with Boolean operators. The keyword and Boolean operator combination used was: (("smart device" OR "voice assistant" OR "smart speaker" OR "IoT" OR "connected toy" OR "smartphone") AND ("covert data capture" OR "covert recording" OR "surveillance" OR "always listening" OR "accidental trigger") AND ("child" OR "children" OR "youth" OR "youths" OR "minors" OR "teenagers") AND ("consent" OR "privacy" OR "data protection" OR "child privacy" OR "youth privacy" OR "ethical implications") AND ("policy" OR "legislation" OR "regulation" OR "PIPEDA" OR

"compliance")). A manual search was also conducted to verify that no relevant papers were missing in the API queries. This search resulted in 4 papers not collected by the Python scripts, which were manually added to our list. The Python scripts used for this process are available in the public GitHub repository at commit e31befc0662511eb7e06194e35e384c04541be23 on the main branch [23]. This process resulted in 1930 academic papers, conference papers, and reputable grey literature (e.g., government and industry reports).

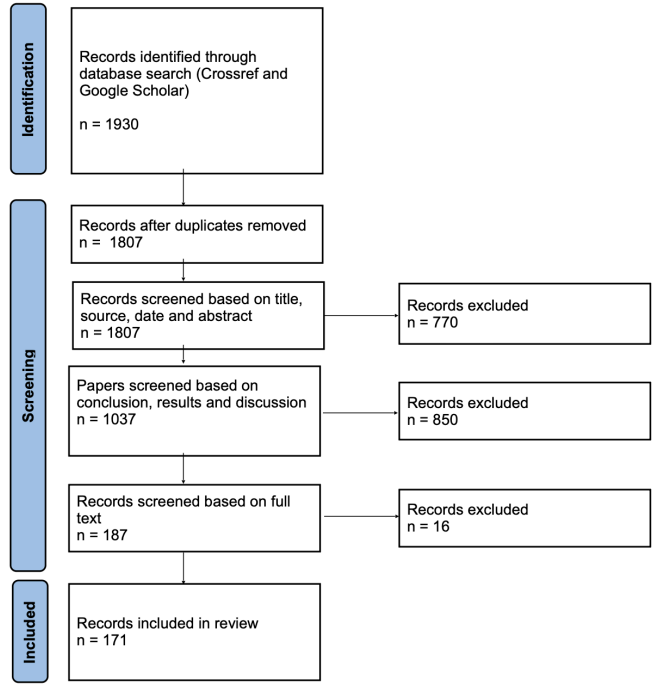

Fig. 1. PRISMA Flow Diagram

### C. Screening and Selection

Following the systematic literature review process, our team began identifying duplicate papers. This resulted in 123 duplicate papers being eliminated from our initial list of 1930 papers, reducing the count to 1807 papers. We then focused on a two-stage process to reduce the collected papers to a relevant selection. A first pass involved considering the title, source, date and abstract of each paper. In addition, the following inclusion criteria were considered:

- Peer-reviewed journal articles, conference proceedings, and reputable grey literature (e.g., government and industry reports) focusing on unauthorized or unintended recordings by smart devices.
- Studies discussing AI-enabled voice recognition, wake-word activation flaws, and real-world incidents of unintentional or non-consensual data capture.
- A global perspective with emphasis on Canadian regulations and privacy frameworks.

An initial review of the papers gathered allowed us to identify works that were not relevant and were subsequently discarded from the list. This process resulted in 770 papers removed from the list, and 1037 papers remaining.

A second pass focused on a more detailed examination of the remaining papers. This was done by considering the conclusion, results and discussion sections of the remaining papers. Each paper was evaluated for relevance to our research question, as well as the aforementioned inclusion criteria. This process resulted in an additional 850 papers being discarded from the review, leaving a remaining 187 papers.

Finally, we began a full-text review of each paper, which resulted in the exclusion of an additional 16 papers, leaving 171 remaining papers included in this study. The complete list of reviewed papers is accessible in the GitHub repository [23].

### D. Extraction of Findings

Having compiled a list of relevant literature, we employed the SCOUR framework to categorize literature into 5 groups, as shown in Table 1. Each paper must be in at least one category, but could be in multiple. To support consistency and efficiency in this process, the research team used OpenAI's ChatGPT (GPT-4, 2025) as an auxiliary tool. GPT-4 was provided with a document outlining the SCOUR framework and was asked to identify the categories each paper was relevant to. The program was also asked to provide a short memo for each paper, outlining the key takeaways and how they are related to the SCOUR framework. The research team evaluated GPT-4's output for each item in the review process, comparing it to their own evaluation of the paper, to ensure accuracy. This process allowed us to group the papers according to their SCOUR categories, as is explored in the discussion section of this paper.

### E. Synthesis of Findings

Using the SCOUR categorization, our team was able to evaluate the distribution of papers associated with each of the SCOUR categories. This enabled us to quickly identify the percentage of papers that aligned with each category. An examination of the literature in each SCOUR category allowed us to identify key themes common across the collected literature. For example, the majority of the literature identified in this study contains some discussion of, or evaluation of, regulatory considerations surrounding smart devices and youth privacy concerns. Less common was literature exploring specific usage and exploitation concerns, an indication that this may be an area well-suited for future research. Further discussion of this evaluation is explored in the subsequent results and discussion sections.

TABLE I. SCOUR CATEGORIES

| SCOUR Category | Description |
|---|---|
| S – Surveillance Mechanisms | How does the device record data? |
| | Does it rely on constant listening, wake-word triggers, or background surveillance? |
| | Are there known vulnerabilities (e.g., misfires, hidden features)? |
| C – Consent and Awareness | How do devices/applications seek (or fail to seek) consent? |
| | How transparent are terms of service about audio/data capture? Is age-appropriate language used? |
| O – Operational Data Flow | Once recorded, where does the data go (cloud, third-party servers, local storage)? |
| | Who has access? How long is it retained, and how is it shared or sold? |
| U – Usage and Exploitation | Are there documented cases of data exploitation for marketing, targeted advertising, or profiling? |
| | Do stakeholders (companies, advertisers, hackers) use recorded data in ways users might not expect? |
| R – Regulatory and Technical Safeguards | What are the existing and proposed technical safeguards and policy measures (PIPEDA amendments) suggested by the authors? |

*F. Paper Writing*

After completing the systematic literature review and exploration of the resulting content under the SCOUR framework, our team began the paper-writing process. This involved summarizing key findings from the included studies and highlighting the most significant themes and insights. This resulted in the identification of existing research gaps, suggesting areas for future research. Throughout this process, we drafted a comprehensive literature review paper, summarizing the key takeaways from this process and providing a foundation for future work in this area.

By employing a systematic review approach, this study aims to understand the risk smart devices pose to the privacy of young users and explore the regulatory, technological, and practical implications of these devices. With this information, we aim to inform the direction of future research in this area, leading to the development of technical and regulatory safeguards for smart devices.

IV. RESULTS

This section summarizes the crucial insights developed throughout the literature review by addressing each of the SCOUR lenses individually. Key takeaways are highlighted for each lens, exploring the study's results and informing the reader about the research questions discussed in the discussion section of this paper.

*A. Surveillance Mechanisms*

Many users frequently interact with smart devices without fully understanding the surveillance mechanisms involved, such as continuous listening or activation through wake-word triggers, leading to heightened concerns about data collection and misuse [24], [25]. These concerns are amplified by reports of hidden features and vulnerabilities that allow background surveillance and inadvertent data recording [26], [27]. Although these technologies offer convenience and safety benefits, especially in family-oriented smart homes, young users and parents alike worry about privacy invasions and unauthorized monitoring of children's activities [28].

As an example of current regulatory limitations, nearly all communications companies evaluated failed to meet the requirements set forth by the Malaysian Personal Data Protection Act [29]. Other studies have found that the always-on, wake-word approach often employed to activate smart devices is prone to inadvertent activation, resulting in covert surveillance of the device's vicinity [16].

To address these anxieties, greater transparency about surveillance practices and user-friendly tools for managing data collection preferences are necessary. Providing clear information on how devices record, use, and protect data can empower young users and families, fostering trust and reducing discomfort with IoT surveillance mechanisms. Educational initiatives that promote awareness of privacy settings and potential risks further support informed decision-making, balancing the convenience of smart technologies with critical privacy safeguards.

Additional research highlights how recent advances in technology have accelerated interest, development and the capabilities of IoT and smart device infrastructure [30]. As smart devices become increasingly prevalent in homes, automobiles and wearable devices, surveillance-enabling devices are increasingly present in the world around us and pose privacy risks to the devices' users and non-users in the vicinity of such devices.

*B. Consent & Awareness*

Young users may struggle with understanding and giving informed consent for data collection practices used by smart technologies, primarily due to complex or unclear terms of service and insufficient age-appropriate language [31]. Similarly, many adult users maintain a poor mental model of how smart devices function and what the privacy implications are [32]. Without an informed understanding, there can be no informed consent, highlighting a need for greater awareness, whether through regulatory means or device manufacturer communication.

Little has been done in a proactive sense to determine exactly what children or teens know about online privacy [33]. Studies show that students don't understand the complex infrastructure of the digital environment and the commercial exploitation of their personal digital data by others [34]. Due to the unseen nature of privacy violations, users' privacy concerns are mainly theoretical, related to the collection and misuse of individual contextual information and their private spaces being invaded by unauthorized individuals and organizations [35].

Studies highlight significant gaps in how devices and applications seek consent, particularly in smart home environments, where transparency regarding data practices is frequently inadequate [24]. Additionally, terms of service are typically not designed for readability, especially for young users, intensifying difficulties in understanding what a user is consenting to, leading to uninformed acceptance [28]. Davis argues that consumers often are unable to make informed decisions about their digital privacy because they are in a position of asymmetric information [36].

To effectively address these issues, clear, age-appropriate communication strategies and user-friendly interfaces are necessary. Educational initiatives aimed at enhancing young users' awareness about privacy rights and data collection practices can empower them to engage more meaningfully in the consent process, building trust and fostering responsible digital citizenship.

*C. Operational Data Flow*

Operational data flows in smart devices often involve complex processes where collected data is transferred and stored across multiple locations, including cloud platforms, third-party servers, and sometimes locally on devices [16], [37]. There is significant concern about who has access to this data and how transparently such practices are communicated to users [25]. Many IoT devices and smart home applications rely heavily on third-party cloud storage, raising critical questions about data retention duration, security, and potential unauthorized access [27].

Moreover, studies indicate that data sharing and selling practices are frequently unclear, exacerbating user uncertainty about privacy risks and potential misuse [28]. Cross-border transfers add another layer of complexity and concern, highlighting substantial compliance and governance gaps, particularly relevant in the context of global IoT deployments [26].

Clear, explicit disclosure of data flow pathways, retention periods, and third-party involvement is essential to mitigate these risks and to enhance trust in IoT ecosystems. Transparent operational practices, robust data management tools, and strict regulatory adherence are necessary to ensure user confidence and safeguard privacy effectively [38].

*D. Usage & Exploitation*

Data collected through smart devices can be exploited for targeted advertising, AI training, surveillance, and in other ways, often unanticipated by end users. Studies show widespread consumer unease about such exploitation, especially when personal information is utilized beyond its original purpose, notably in marketing or profiling without explicit user knowledge [39]. Research also highlights how smart home ecosystems, while intended for convenience and safety, often become hubs for targeted data extraction practices, creating vulnerabilities and opportunities for unethical exploitation [40].

Additionally, there are concerns about unethical data use, including manipulative practices or unauthorized profiling, specifically targeting vulnerable populations such as children and youth. These scenarios not only demonstrate potential breaches of ethical standards but also underline the tangible impacts of data misuse, emphasizing the urgent need for stronger safeguards and transparent data-use policies to protect minors from exploitation in digitally connected environments.

However, while there are some documented cases of inadvertent recording and data exploitation [16], [41], the existing research is limited when examining actual instances of inadvertent recording and data use and misuse by device manufacturers and third parties. With the increasing prevalence of smart devices, this represents a critical research gap. A thorough understanding of current data practices and instances of exploitation is critical to informing regulatory and ethical guidelines for the future of AI and smart device development.

*E. Regulatory & Technical Safeguards*

Strengthening regulatory and technical safeguards to protect user privacy in smart technologies has become increasingly important, particularly as it relates to Canada's evolving privacy landscape. Researchers advocate strongly for technical improvements such as advanced encryption, local data processing, and enhanced wake-word detection to mitigate risks related to unauthorized access and unintended data capture [42]. Typically, IoT and smart devices in a user's home will transmit data to remote cloud servers, where the data is processed and stored [37]. The data flow steps of transmitting to remote servers and storing on those servers can expose privacy-sensitive data, such as audio recordings, to an increased risk of cyberattacks. In other cases, such as smart toys, data might first be transmitted to an application on a smartphone before being transmitted to a cloud-based service for processing and storing the data, presenting additional attack vectors for malicious actors [43].

The regulation of data protection is also a primary economic concern, particularly as it influences the practices of information-intensive businesses in this modern era [44]. Additionally, policy-oriented studies suggest that Canada's privacy regulations, especially the PIPEDA, should incorporate key aspects of standards like the GDPR of the EU. Recommendations include incorporating clearer consent processes, stricter guidelines on data retention, and increased transparency regarding third-party data usage [45]. An alternative, or perhaps additional, approach includes a privacy-preserving system based on differential privacy mechanisms, which has also been explored [46].

Despite these recommendations, significant gaps persist between proposed measures and actual implementation, highlighting the need for improved oversight, frequent compliance assessments, and user-focused education [47]. In addition, the modular and continually developing nature of the IoT ecosystem presents various attack vectors for malicious actors. Modern, ongoing analysis of the IoT data lifecycle, architecture, and stakeholders should inform data privacy and security regulation [48], [49], [50].

As with any new technology, it is essential to carefully consider the implications and develop appropriate regulations and guidelines to protect privacy and individual rights [51]. Bridging the existing gaps through comprehensive policy enhancements, combined with technological approaches to privacy preservation, will significantly reinforce privacy protections and build user trust in IoT ecosystems [52].

Beyond these considerations, the synthesis of findings across the reviewed studies also points to several recurring behavioral and perceptual dimensions of privacy. While the SCOUR framework was employed to organize and classify the reviewed studies, the analysis further revealed other recurring thematic dimensions such as privacy risks, data-sharing benefits, algorithmic trust and transparency, user confidence in privacy management, and privacy-protective behaviors.

## V. Discussion

The results from our systematic literature review provide a detailed and comprehensive understanding of how smart devices may covertly capture private conversations and their implications for youth privacy. For this review, only papers that specifically addressed one or more of the themes within the SCOUR framework were selected. This included both user studies and general discussions or opinion papers, as long as they fit at least one of the themes. A pie chart in Fig. 2 illustrates the distribution of papers discussing these themes, including both the number of studies and their proportional representation across the five SCOUR categories. It is important to note that most papers fit into multiple categories; a total of 631 SCOUR tags were applied across the 187 included studies. Each research question outlined in Section III is addressed through thematic analysis, with subsequent subsections providing insights and answers to each question based on the thematic findings.

*A. Can smart/AI-enabled devices, applications, intentionally and surreptitiously record private conversations? If so, what evidence exists of this practice?*

This is the primary research question which guided the literature review on the implications of smart/AI-enabled device voice eavesdropping.

Research reveals growing evidence that smart and AI-enabled devices can, and in some documented cases do, surreptitiously record private conversations. This functionality is often tied to voice-activated assistants like Alexa, Siri, and Google Assistant, which rely on passive listening to detect wake words. However, concerns arise when these devices capture audio without explicit user consent or outside the intended trigger. For example, Sweeney and Davis explore how smart voice assistants blur the boundary between public and private spaces, identifying scenarios where these devices listen beyond their intended scope, raising critical questions about surveillance and consent [25]. Similarly, Elvy discusses how commercial law and user agreements often mask intrusive surveillance practices under broad terms of service, allowing companies to gather vast amounts of conversational data without transparent user knowledge [53].

Research on "Modern Privacy Threats and Privacy Preservation Techniques in Smart Homes" underscores that always-on microphones in smart home ecosystems are inherently vulnerable to unauthorized access and may be exploited by malicious actors or used by companies for targeted advertising or selling to third parties [54]. Additionally, there exists technical research exploring how smart devices are built. One paper evaluates hardware-based voice detection systems and their energy efficiency, which notes how these systems are optimized for continual background listening, a feature that, while practical, increases the risk of unintended recordings [55].

As the regulatory landscape continues to evolve, high-profile legal decisions have played a significant role in shaping organizational behavior. One such case "has catalyzed significant changes in consent practices, reshaping the digital advertising ecosystem and compelling businesses to reassess their data protection strategies" [56]. Another paper states that "legislations, however, must not be viewed as impediments to business but as business enablers that ensure successful conduct of business while balancing the rights of the individuals…" [57]. Collectively, the literature indicates that both the policies and architecture of smart devices contribute to the potential of surreptitious recording. While not all devices are designed to conduct unethical or malicious recording, the capacity to do so may still exist, and the lack of transparency and regulation amplifies concerns.

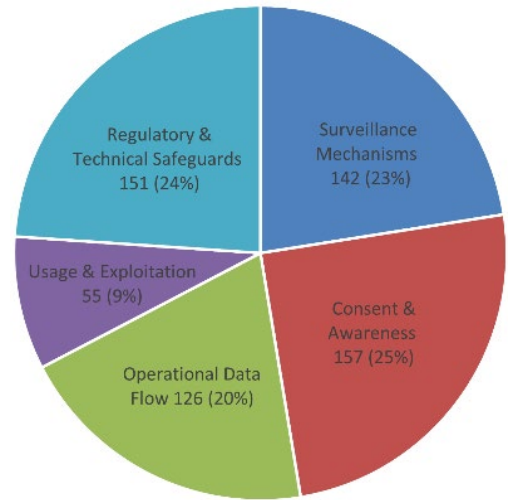

Fig. 2. Breakdown of SCOUR Categories

*B. What mechanisms allow such recordings to occur?*

Smart and AI-enabled devices such as IPAs or VAs utilize a range of technical mechanisms that enable the surreptitious recording of conversations, including microphone arrays and machine learning voice-recognition technology. IPAs or VAs can be found in various devices, including smart speakers, such as Amazon's Alexa, cell phones, such as the Siri assistant in Apple's iPhone, and even in televisions, automobiles and children's toys [14], [15], [32], [58], [59]. Research has found that IPAs are primarily used for tasks such as playing music, answering questions, controlling IoT and home automation devices, setting timers and alarms, checking the weather, etc. [16].

To conduct these activities, these devices rely on always-on microphones, wake-word detection systems, and embedded sensors that passively monitor the environment for activation cues. This architecture enables passive surveillance by default, even if recordings are only meant to begin after activation.

Prior research has indicated that smart devices are prone to 'waking' and hence processing the audio in their vicinity, without a wake word having been invoked intentionally by a user [16]. This presents a risk of otherwise consensual device users being recorded inadvertently, at moments when they are unaware a recording is taking place, hence violating their consent and privacy expectations.

Furthermore, devices used knowingly by a consensual user may still pose ethical concerns. There is a risk of a smart device being used knowingly by a user, also capturing the audible conversations of other, non-consenting users in the vicinity of the device. As IPAs are increasingly used in portable devices such as cell phones, the risk of surreptitious recording of a device's ambient environment is increasing.

In addition to accidental recording, smart devices are subject to various attack vectors. Attackers may seek to conduct unauthorized surveillance recordings through attacks such as voice squatting, so-called evil maid attacks, or the use of inaudible commands being transmitted to a device [60], [61]. Malicious actors may also seek to compromise the legitimate data flow of the smart device to intercept consensual recordings made by the user [15], [62].

*C. How is the collected data used, stored, and shared?*

Smart devices typically transmit voice data collected to cloud servers, where they undergo processing, storage, and integration with other datasets [16]. This process often occurs without users fully understanding where the data goes or how long it is kept [16]. In some cases, audio recordings are stored on servers by the service provider. This exposes such data to cybersecurity threats, as demonstrated in the 2017 cyberattack on the CloudPets company in which approximately two million voice recordings between children and parents were exposed to hackers [15], [62].

Interestingly, a 2024 study of security practices in various smart devices found that despite having the hardware capability for security best practices, such as secure boot or encryption, many smart devices neglected to actually use these privacy-protecting technologies [18].

Beyond functional use, collected voice data is often exploited for secondary purposes such as personalized advertising, behavior profiling, and even resale to third parties. However, the research exploring the long-term use of collected data is limited. Most companies provide broad terms of service, but limited information on the exact uses of collected data. Understanding the use, storage and sharing of collected user data is critical to informing adequate policy approaches, and future research in this area is strongly indicated.

*D. Are there documented cases where companies or third parties have leveraged such recordings?*

There are documented cases where companies and third parties have leveraged recordings from smart or AI-enabled devices. One notable example comes from studies examining the use of smart home technologies, where researchers found instances of audio recordings being reviewed by human contractors for quality assurance. This was often without users' explicit knowledge or consent [63]. Such reviews were conducted by third-party workers who had access to sensitive and private conversations. Additionally, broader surveillance concerns have been raised in the context of AI development, where massive datasets, including audible voice and conversation data, are collected, processed, and shared across organizations, often lacking transparency or accountability [64].

While existing research and specific case studies in this area are limited, the existing work demonstrates that usage and exploitation of voice recordings are more than theoretical risks and have already occurred in various forms. This reflects a significant risk to the user's privacy and highlights the significant challenges faced in the regulation and enforcement of policy regarding smart technology.

*E. What are the technical, policy, and legal limitations regarding surreptitious recording?*

Within Canada, PIPEDA provides a broad approach to the regulation of data collection and use by corporate entities [65]. This is supplemented at the provincial level by various regulations. For example, British Columbia has the Freedom of Information and Protection of Privacy Act (FIPPA), which provides privacy regulation applicable to public sector entities, and the Personal Information Protection Act (PIPA), which regulates private sector entities [66], [67].

With the dominance of international corporations such as Amazon and Apple in the smart device ecosystem, there is literature that suggests a need to develop an appropriate infrastructure to enhance the security of the users of IoT devices, which is not necessarily specific to a specific nation [68].

Regarding corporate entities, internal company guidelines vary significantly. While some companies have introduced privacy dashboards or opt-out features, the adoption of such features is inconsistent [16].

Internationally, there are robust privacy frameworks that can be used to inform further development and amendments to the existing Canadian regulations. In the European Union, the GDPR provides extensive privacy oversight, including privacy regulations specifically focused on protecting the privacy rights of children [69].

As AI tools such as smart speakers become more prevalent, Canadian regulators should consider an ongoing approach to privacy regulation to keep pace with rapid technological developments in the field. The European Union's AI Act is an example of a regulatory approach that can inform future policy in this area [70]. On a global level, research indicates that there is an advantage to international collaboration on data privacy, suggesting that despite advances in domestic regulation, Canada should engage in a broader discussion and work towards a robust global privacy ecosystem [71].

*F. What preventive measures can individuals and regulators take to mitigate unintended recordings?*

To reduce the risk of unintended recordings by smart and AI-enabled devices, both individual users and regulatory bodies can take proactive steps. On the individual level, users can make use of privacy settings such as disabling microphone access when not needed and opting out of data-sharing programs when such options are available. Additionally, increasing user awareness through transparency reports and interface design improvements can empower users to make informed decisions about their privacy [25].

However, studies have shown that users generally exhibit confusion around when speakers are recording and what information is captured, and users typically have a poor mental model of smart speaker technology [32], [61].

Some technological steps can be taken by users to protect and reinforce their privacy. For example, Amazon and Google's smart speakers are equipped with a user accessible logging functionality. This provides users with access to transcripts of audio conversations and the actual recordings and provides users with a mechanism for reviewing and deleting log entries. However, research shows that a sizable portion of users were unaware of such features, and among those who were aware that the feature existed, many had never utilized it [16].

Broadly speaking, countries should adopt a personal data protection law, and those that already have a law should continually refine it further [72]. Furthermore, technical solutions exist at the individual, developer and regulatory levels. For example, most wake-word-activated smart devices employ voice recognition for identification purposes, but without providing authentication [73]. Identification allows a smart device to tailor their responses to a specific individual, but does not prevent unauthorized users from accessing and interacting with the device.

Without appropriate authentication protocols, a smart device can be activated inadvertently or by malicious actors, leading to surveillance of the device's surroundings unbeknownst to the owner or others in the vicinity of the device. Possible authentication solutions include biometric authentication of a user's identity, such as facial recognition. This would require the addition of cameras to devices such as Amazon's Alexa, as well as the consensual opt-in of the user. It is worth noting that the incorporation of a camera into smart devices poses additional privacy and security risks, such as the possibility of inadvertent video recording.

*G. Future Works*

The synthesis of findings through the SCOUR framework revealed recurring dimensions underlying user interactions with AI-enabled technologies. While existing studies emphasize fragmented aspects of privacy, a closer thematic integration points toward foundational constructs that consistently shape privacy-related behavior and perceptions, including perceived privacy risks, perceived privacy benefits, algorithmic trust and transparency, privacy self-efficacy, and privacy-protective behavior. These constructs collectively capture the balance between users' awareness, trust, and agency in managing their personal data. They also provide a structured lens for quantitatively examining how individuals, particularly young digital citizens, evaluate and respond to privacy concerns in AI-driven contexts. Accordingly, these constructs form the conceptual basis for future research aimed at empirically investigating how privacy perceptions and behaviors manifest among youth, using structural equation modeling on survey data to uncover relationships and inform the design of privacy-aware smart systems.

*H. Limitations*

This study has several limitations. The literature review was restricted to English-language articles, which may limit the generalizability of the findings. Furthermore, the rapid pace of smart device and AI development means that the literature may not fully cover all the latest practices, suggesting a continual need for updated research to address ongoing developments in privacy and smart technology. Research has shown that it is challenging to design a real-time privacy protection mechanism [74]. This study reflects the current state of research, which is and should be viewed as a snapshot in time rather than a definitive account of the field.

## VI. Conclusion

This paper highlights the growing privacy risks posed by smart and AI-enabled devices for both adult and youth users. Using the SCOUR framework, we found that many of these devices covertly collect data through always-on features, often without clear user awareness or consent. We identified recurring issues in surveillance practices, data handling, and regulatory gaps, as well as a lack of transparency in how personal data is collected and used. While existing regulations like PIPEDA offer some protection, enforcement and public understanding remain limited. To address this, stronger safeguards, clearer communication, and youth education are essential. Specifically, studies have found that devices themselves provide inadequate information regarding their data collection policies. Regulation that requires device manufacturers to clearly outline the specific data collection, storage, and usage limitations is a clear area for regulatory improvement, providing agency to users and contributing to informed consent regarding smart devices. To build a future of trust in smart technologies, there is a critical need for user-centric design that emphasizes transparency, ethical data practices, and clear communication, especially for youth and their guardians. Educational initiatives must play a central role in raising awareness, and policies must evolve to reflect emerging risks and technological advancements. Protecting digital privacy is not just a technical challenge; it's a societal responsibility that demands thoughtful design, inclusive policy, and a deep respect for user autonomy.

Acknowledgment

This project has been funded by the Office of the Privacy Commissioner of Canada (OPC); the views expressed herein are those of the authors and do not necessarily reflect those of the OPC.